\documentclass[aps,prl,twocolumn,superscriptaddress,showpacs]{revtex4-1}
\usepackage{graphicx}
\newcommand{\mSR}{$\mu$SR }
\newcommand{\Tc}{$T_{c}$ }
\newcommand{\bNMR}{$\beta$-NMR }
\newcommand{\Li}{$^8$Li }
\newcommand{\Liplus}{$^8$Li$^+$ }

\begin{document}


\title{Observation of slow order parameter fluctuations in superconducting films using \\
$\beta$-detected NMR}


\author{E.~Morenzoni}
\email[E-Mail:]{Elvezio.Morenzoni@psi.ch} \affiliation{Paul Scherrer
Institut, Laboratory for Muon Spin Spectroscopy, 5232 Villigen PSI,
Switzerland}
\author{H.~Saadaoui}
\affiliation{Paul Scherrer
Institut, Laboratory for Muon Spin Spectroscopy, 5232 Villigen PSI,
Switzerland}
\author{D. Wang}
\affiliation{Department of Physics and Astronomy, University of British Columbia, Vancouver, BC, Canada, V6T}
\author{M. Horisberger}
\affiliation{Paul Scherrer Institut, Laboratory for Developments and Methods, 5232 Villigen PSI,
Switzerland}
\author{E. Kirk}
\affiliation{Paul Scherrer
Institut, Labor f\"ur Mikro- und Nanotechnologie, CH-5232 Villigen PSI,
Switzerland}
\author{W.A. MacFarlane}
\affiliation{Chemistry Department, University of British Columbia, Vancouver, BC, Canada, V6T 1Z1}
\author{G. D. Morris}
\affiliation{TRIUMF, 4004 Wesbrook Mall, Vancouver, BC, Canada, V6T 2A3}
\author{K.H. Chow}
\affiliation{Department of Physics, University of Alberta, Edmonton, AB, Canada, T6G 2G7}
\author{M.D. Hossain}
\author{C.P. Levy}
\affiliation{Department of Physics and Astronomy, University of British Columbia, Vancouver, BC, Canada, V6T}
\author{T.J. Parolin}
\affiliation{Chemistry Department, University of British Columbia, Vancouver, BC, Canada, V6T 1Z1}
\author{M.R. Pearson}
\author{Q. Song}
\affiliation{Department of Physics and Astronomy, University of British Columbia, Vancouver, BC, Canada, V6T}
\author{R. F. Kiefl}
\affiliation{Department of Physics and Astronomy, University of British Columbia, Vancouver, BC, Canada, V6T}
\affiliation{TRIUMF, 4004 Wesbrook Mall, Vancouver, BC, Canada, V6T 2A3}

\date{\today}

\begin{abstract}
We report \bNMR investigations of polarized \Li implanted in thin Pb and Ag/Nb
films. At the critical superconducting temperature, we observe a singular peak in the spin relaxation rate in small longitudinal magnetic fields, which is attributed to fluctuations in the superconducting order parameter. However, the peak is more than an order of magnitude larger than the prediction based on the enhancement of the dynamic electron spin susceptibility by superconducting fluctuations
and reflects the presence of unexpected slow fluctuations. Furthermore the fluctuations are rapidly suppressed in a small magnetic field, which may explain why they have not been observed previously with conventional NMR or NQR.
\end{abstract}

\pacs{74.25.nj, 74.40.-n, 74.78.-w}


\maketitle

%
%
Near the superconducting critical temperature \Tc, thermodynamic fluctuations of the order parameter occur, leading to the appearance of pairing correlations with critically
increasing life time on approaching $T_c$ from above \cite{Larkin09}. These fluctuations are easier to see above $T_c$ in the form of excess conductivity (paraconductivity), tunnel conductivity, enhanced diamagnetism, and in
many other electronic properties \cite{Skocpol75}.
The importance of fluctuations increases in systems of lower dimensionality (on the scale of the Ginzburg-Landau coherence length) such as thin films, wires and small particles.
Unconventional superconductors such as cuprates, with their quasi 2D electronic structure, low superfluid density and short coherence length,
are even more susceptible to fluctuations \cite{Emery95}. In these systems phase fluctuations are thought to be largely responsible for the pseudogap regime and may be connected to the presence of a quantum critical point \cite{Orenstein00}.

NMR can give detailed information on the normal and superconducting electronic state and has played a pivotal role in elucidating the nature of low- and high-$T_c$ superconductivity \cite{Rigamonti98}.
In low-\Tc superconductors the observation of the Hebel-Slichter peak was a profound and unambiguous confirmation of BCS theory. Also, in high-\Tc superconductors (HTS) NMR measurements of the local spin susceptibility first revealed the unusual pseudogap state \cite{Alloul89}.
The nuclear spin relaxes due to time dependent fluctuations of the local magnetic field, whose transverse components induce transitions between the nuclear spin levels.
In a metal the most important contribution is from spin fluctuations of the conduction electrons which are coupled to the nuclear spins though the magnetic hyperfine interaction. This nuclear spin relaxation rate can be expressed in terms of the dynamical electron spin susceptibility
$\chi''$ and gives rise in the normal state of a non interacting Fermi gas to the well known Korringa law $\frac{1}{T_1T}=const$ of the longitudinal spin relaxation rate $\frac{1}{T_1}$.
The effect of superconducting fluctuations on $\chi''$ are generally calculated, using diagrammatic techniques, to different levels of approximation (see \cite{Larkin09} for an overview).
In an s-wave superconductor without strong pair breaking effects one expects a singular peak in $\frac{1}{T_1}$ at $T_c$ in a very narrow temperature range $\epsilon=\frac{T-T_c}{T_c}$.
However, until now no singular behavior of the spin-lattice relaxation at $T_c$ has been detected, either in conventional or in unconventional superconductors.
In conventional superconductors the observation of the effect of superconducting fluctuations on NMR spectra and relaxation is believed to be difficult due to the smallness of the effect
on the order of the Ginzburg-Levanjuk number $G_{i}$ ($ \approx \frac{k_B T_c}{E_F}$) \cite{MacLaughlin76}. For d-wave pairing the dominant contribution, the so-called Maki-Thomson term, is suppressed \cite{Kuboki87,Maki68}. In HTS, comparison with theory is further complicated by the delicate interplay of contributions with opposite sign and magnetic field dependence.
Moreover, the presence of antiferromagnetic fluctuations and the unusual normal state background make it difficult to extract the fluctuation contributions, leading to controversial interpretations of the experimental data, which instead of a critical enhancement show a suppression of $\frac{1}{T_1}$ at \Tc \cite{Carretta96,Mitrovic99}.

In this letter, we report low frequency  $\beta$-NMR measurements in thin films (single and NS bilayers) of low $T_c$ superconductors. We measured RF resonance spectra and the longitudinal spin relaxation rate $\frac{1}{T_1}$ of $^8$Li implanted in a thin Pb film and in the Ag layer of several Ag/Nb bilayers, where superconductivity is induced by proximity effect. In both films we observe  a large diamagnetic shift in the resonance frequency below $T_c$ as expected from Meissner screening of  the  applied field. However, there is also a well defined  peak in  $\frac{1}{T_1}$ in a narrow temperature range near the critical temperature $T_c$, which is clearly due to fluctuations of the superconducting order parameter. The fluctuations are much slower, and  the region around $T_c$ where they are observed is larger than predicted from current theory of Gaussian fluctuations. Furthermore there is a strong field dependence which may explain why these fluctuations have not been reported previously with conventional NMR.
This experiment shows that it is possible to observe superconducting fluctuations by the NMR technique and paves the way to investigations in other classes of superconductors (e.g. HTS), low dimensional systems
and mesoscopic devices (e.g. Josephson arrays).

The $\beta$-NMR experiment was carried out at TRIUMF using a 28 keV beam of \Liplus  produced at the isotope separator and accelerator facility (ISAC). A large nuclear polarization (70\%) is generated in flight using a collinear optical pumping method \cite{Morris04}. The hyperpolarized beam had  a beam spot of about 3 mm diameter and a typical rate of $10^{6}$  $^8$Li/s.  The  samples were mounted on a cold finger cryostat contained in the UHV chamber of the low field $\beta$-NMR spectrometer. The spin polarization was perpendicular to the beam direction but parallel to both the sample surface and a small external magnetic field B$_0$ between 0.7 and 3 mT.
In $\beta$-NMR the nuclear spin polarization is monitored through the anisotropic $\beta$  decay of \Li (S = 2, gyromagnetic ratio 6.3015 MHz/T, electric quadrupole moment Q=+ 31.4 mb) which has a mean lifetime
$\tau$  = 1.2 s. The emitted beta has an average energy of about 6 MeV, which allows it to pass easily through stainless steel windows in the ultra high vacuum (UHV) chamber. The beam energy can be chosen between 1 and 28 keV \cite{Morris04}.

The samples were RF-sputtered onto epitaxially polished quartz and sapphire substrates using research grade materials. Resistivity measurements give for Pb (260 nm) $T_c$= 7.12 K, $\Delta T_{c} \lesssim $ 8 mK with
mean free path $\ell \approx 48$ nm which is less than the BCS coherence length  $< \xi_0 \cong 83$ nm. Also since $k_BT_c \hbar^{-1}\tau = 0.03 \ll 1$, where $\tau$ is electron scattering time the sample is in the dirty limit.
For Ag(29 nm)/Nb(252 nm), the critical temperature of the entire bilayer $T_c$= 9.15 K, $\Delta T_{c} \lesssim $ 40 mK, $\ell_{Ag} \approx 40$ nm and $\ell_{Nb} \approx 250$ nm.

\begin{figure}[h]
    \centering
 \includegraphics[width=1.0\linewidth]{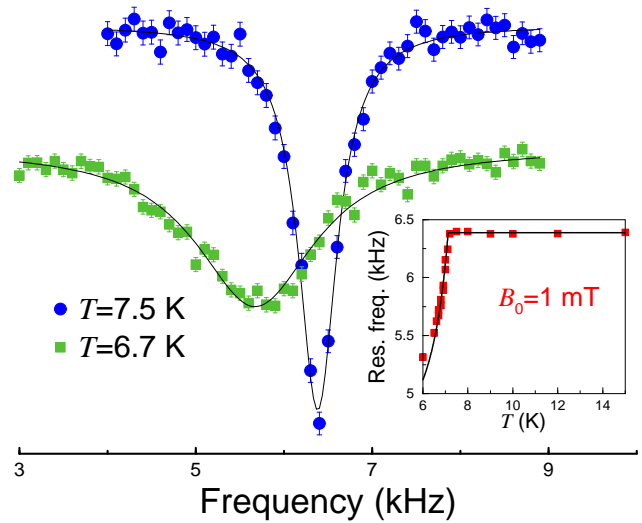}
  \caption{$\beta$-NMR resonance spectrum above and below T$_{c}$ for $^8$Li stopped in the Pb film. The insert shows the temperature dependence of the resonance frequency. The curve is a fit to the London model.}
  \label{Pb_resonance}
\end{figure}

Fig. \ref{Pb_resonance} shows resonance curves of \Liplus implanted at 5 keV in the Pb sample with a magnetic field of 1 mT applied parallel to the surface of the film. At this energy the implantation range of \Liplus  extends to $\sim$ 100 nm with a mean implantation depth and rms of 34.05 nm and 19.22 nm, respectively.
The temperature dependence of the resonance frequency is an unambiguous signature of the Meissner state (inset Fig.  \ref{Pb_resonance}). The frequency shifts data fits well to a solution of the London equation taking into account the implantation profile yielding an effective penetration depth
$\lambda_{eff}(T)$=$\frac{\lambda_{L}(0)}{\sqrt{1-(\frac{T}{T_c})^4}} \sqrt{1+\frac{\xi_{0}}{\ell}}$ with $\lambda_{eff}(0)$= 62.2$ \pm$ 0.3 nm in good agreement with a LE-\mSR measurements on this sample \cite{Suter05} and literature values \cite{Gasparovic70}. A sharp decrease in the resonance frequency in the Ag/Nb film was observed   below $T_c$   at an implantation energy of 4keV. In this case the signal originates entirely from the Ag overlayer of the Ag/Nb bilayer.
Knight shift $\beta$-NMR measurements in elemental metals such as Ag and Au indicate that below $\sim$ 100 K the Li is static and occupies the interstitial octahedral site, which is favored over the smaller tetrahedral site \cite{Morris04,Parolin08,Hossain09}. Since Pb has the same fcc structure it is therefore reasonable to assume that Li is mainly in the octahedral interstitial site over the temperature range of our experiment.

\begin{figure}[h]
  \includegraphics[height=1.0\linewidth,angle=-90]{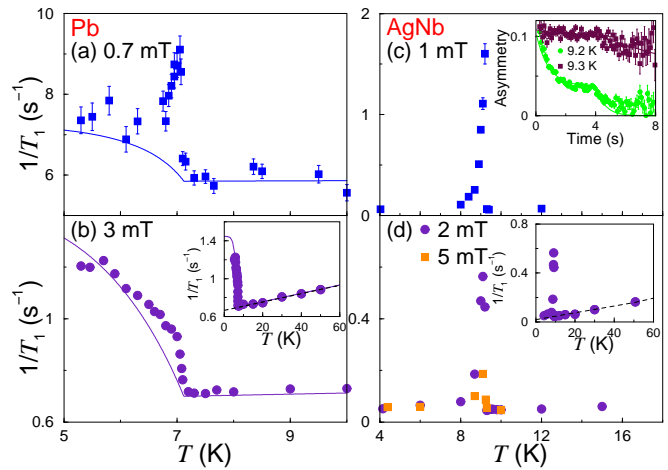}
  \caption{Spin lattice relaxation rate $\frac{1}{T_1}$ versus temperature
from single  exponential fits to relaxation data. (a) and (b): Pb film (260 nm) $^8Li$ implantation energy 5 keV. (c) and (d): Ag(29nm)Nb(252nm) bilayer $^8Li$ implantation energy 4 keV.
Insets (c): typical polarization spectrum; (b) and (d): extended temperature range showing the Korringa contribution. The solid curves show the contribution from the host nuclear dipole dynamics in Pb.}
\label{Pb_AgNb_lambda}
\end{figure}

\
Spin relaxation measurements were carried out with a pulsed beam of length 1-4 s with the RF field off.
The insert of Fig. \ref{Pb_AgNb_lambda}(c) shows the time evolution of the asymmetry signal of the \Li stopped in Ag. The decay of the signal is a direct measure of $\frac {1}{T_1}$.
Fitting the time dependent asymmetry during and after the pulse, by taking into account the constant implantation rate during the pulse and the exponential relaxation, we obtain  $\frac{1}{T_1}$.

Fig. \ref{Pb_AgNb_lambda} plots the temperature dependence of $\frac {1}{T_1}$ for both samples.
Note in particular the sharp increase at $T_c$ within $\lesssim $ 50 mK followed by a smooth decrease
below T$_c$. Such  a peak at T$_c$  is unambiguous evidence  for critical slowing down  of the fluctuations  in the superconducting order parameter. Before discussing this  in more detail it is   important to first  understand the data in the normal state. Above T$_c$ the relaxation rate increases linearly with T as expected from Korringa relaxation  in the normal state (inset in Fig. \ref{Pb_AgNb_lambda}(b),(d)).
Extensive $\beta$-NMR relaxation studies in pure elemental metals such as Ag, Au and Nb have shown that in the normal state at sufficiently high field $T_1$ is dominated by the magnetic field-independent, temperature-linear Korringa relaxation due to the direct hyperfine coupling between the \Li  and  the conduction electrons. The measured slopes 0.0028(1) $(sK)^{-1}$ for Ag and 0.0044 $(sK)^{-1}$ for Pb,  are close to those found in other elemental metals \cite{Hossain09}.

Spin  relaxation for \Li in  Pb at  low magnetic fields is  more complicated due to cross  relaxation  between the host Pb spins and \Li.  Thus in addition to the direct  Korringa relaxation there is also an indirect  contribution   from the  Korringa relaxation  of the host $^{207}$Pb   spins (S=1/2 and  22\% abundant) which are coupled to the \Li through  a magnetic dipolar interaction.
The interaction produces a significant cross relaxation between \Li and $^{207}$Pb  nuclear spins in low field where the Larmor frequencies of the $^{207}$Pb and \Li are small and thus nearly equal.
As shown in \cite{Hossain09b} this process characteristically gives rise to a strong Lorentzian field dependent $\frac {1}{T_1}$ evident in the normal state in Fig. \ref{Pb_AgNb_lambda}(a),(b).
Consequently, the reduction of the local field in the superconducting state leads to a temperature dependent enhancement of $\frac {1}{T_1}$ over the normal state values starting at \Tc and extending to lower temperatures. We have calculated this contribution fitting the field dependence of the relaxation rate above \Tc to determine magnitude ($\Delta$B = 0.01 mT)  and fluctuation time ($\tau_c$ = 3 10$^{-5}$ s) of the $^{207}$Pb dipolar fields. Below \Tc we take into account the reduction of the local field acting on the \Li nuclei due to the Meissner effect with penetration depth $\lambda_{eff}(T)$ as determined above from the resonance data.
The solid curves in Fig. \ref{Pb_AgNb_lambda}(a),(b) show the results of the calculation including a small Hebel-Slichter contribution, due to opening of the gap and coherence effects, which we calculated following the theory of \cite{Fibich65}.
It is clear that the remaining singular peak observed at $T_c$ cannot be due to due to the Hebel-Slichter effect because the latter appears as a much broader field-independent peak at $T \sim 0.85 T_c$. Furthermore the Hebel-Slichter peak represents a barely factor two enhancement over the normal state Korringa value of
0.0313 $s^{-1}$ for Pb (see also below).
The peak we observe occurs exactly at $T_c$ as shown in Fig. \ref{Pb_AgNb_lambda} and is easily suppressed in a few mT external field. This is clear evidence that the peak is due to critical fluctuations
in the superconducting order parameter.

In terms of the mean field Ginzburg-Landau order parameter $\Psi$, the superconducting fluctuations are related to~  \mbox{$\langle~|\Psi|^2\rangle\neq0$} although $\langle\Psi\rangle=0$. This state of local phase coherence can be represented by the presence of superconducting droplets of size $\xi(T)=0.85 \sqrt{\xi_0 \ell} \sqrt{\frac{1}{\epsilon}}$
and life time of the order of
$\tau_{GL}(T)=\frac{\pi \hbar}{8 k_B T_c \epsilon}$ \cite{Larkin09}.
The non-equilibrium pair density $n \propto \langle|\Psi|^2\rangle$ leads to fluctuation induced diamagnetic susceptibilty and paraconductivity.
The effect of superconducting fluctuations on physical observables is generally
calculated in the Gaussian regime by the diagrammatic technique in a microscopic theory involving the hyperfine contact coupling.
The acceleration of Cooper pairs of limited lifetime contributes directly to the conductivity with the so-called Aslamasov-Larkin (AL) term \cite{Aslamazov68}.
An additional indirect contribution (the so-called Maki-Thomson term, MT \cite{Skocpol75}) reflects the conductivity increase induced by fluctuations decaying into quasiparticle pairs of nearly opposite momenta that maintain their correlation and continue to be accelerated as though they were paired \cite{Larkin09}.
Other contributions are given by the diagrams due to the reduction of the single-particle density of states by the superconducting fluctuations (DOS).

For NMR observables, the interplay and importance of these contributions is different than in the case of conductivity.
Due to spin-singlet pairing the direct Cooper pair AL contribution to the (static and dynamic) spin susceptibility is absent and
the anomalous MT term is the only positive and dominating contribution to the relaxation rate.
A physical picture of the MT relaxation process is the following. When the electron is scattered by the nucleus changing its spin orientation and
momentum to the opposite value, it can pass again the previous trajectory moving in the opposite direction.
Due to the retarded character of the pairing interaction, a Cooper pair can form and this turns out to be a
new spin relaxation mechanism.

However, the phenomenon reported here cannot be explained by the MT enhancement of the local spin susceptibility because i) the magnitude of the peak with respect to the
Korringa relaxation is much larger ii) it appears in a wider temperature range iii) and because of its strong field dependence.
In the 2D limit, which is well fulfilled in the temperature interval where we observe the jump in spin-lattice relaxation ($\xi(\epsilon) \gtrsim d$ for $\epsilon \lesssim 3 \cdot 10^{-2}$),
and dirty case, the MT enhancement of the NMR relaxation rate over the normal Korringa value can be analytically expressed in the case of static ($\omega \rightarrow 0$) long wave length fluctuations to be \cite{Kuboki87,Randeria94}:
\begin{eqnarray}
\frac{\frac{1}{T_1}^{MT}}{\frac{1}{T_1}^{n}}=\frac{\pi\hbar}{8E_F\tau}\frac{1}{\epsilon-\gamma_{\Phi}}\ln(\frac{\epsilon}{\gamma_{\Phi}})
\label{MTformula}.
\end{eqnarray}
\
which diverges logarithmically for $\epsilon \rightarrow 0$.

The size of this contribution is very sensitive to pair breaking processes expressed by the pair breaking parameter $\gamma_{\Phi}$.
Taking for a conservative estimate for Pb only the thermal phonon contribution $\gamma_{\Phi} \approx \frac{\pi^2}{4}\frac{(T_c)^2}{\theta_D^2}=1.57 \cdot 10^{-2}$,
and $\epsilon = 7 \cdot 10^{-4}$ (which corresponds to the temperature stability of our experiment, 5 mK), we
obtain an estimate for the jump in the relaxation rate at $T_c$ which is more than an order of magnitude smaller than observed.
For the Ag/Nb film the calculation underestimates the peak by two orders of magnitude.
The discrepancy is larger if one considers that additional effects not included in Eq. (\ref{MTformula}), such as strong coupling ($\lambda_{e-ph}^{Pb}=1.547$) \cite{Eschrig00} and some degree of three dimensionality \cite{Kuboki87} further reduce the magnitude of the MT term.
Also the existing theory does not account for the strong field dependence of the peak which is suppressed by a few mT.
Such field dependence is not expected for the MT processes of the dynamical spin susceptibility \cite{Fay01,Eschrig99}.

Although more quantitative estimates must await further investigations
the results indicate that the singular fluctuation contribution to the dynamic spin susceptibility as calculated in lower order cannot predict magnitude and frequency spectrum of the effect.
Other mechanisms, that add to the direct contact interaction have to be considered. Since the order parameter is complex, phase fluctuations can generate fluctuating supercurrents and magnetic fields relaxing the spin probe even in cases where the contact term is small or zero \cite{Cyrot72}.
Local fluctuating fields are also associated to diamagnetic fluctuations, which manifest themselves in the diamagnetic susceptibility above $T_c$.
The suppression of  $\frac{1}{T_1}$ as a function of magnetic field  indicates that there is a decrease in the size  (and corresponding correlation time) of the fluctuating domains with increasing applied field, as observed
in the field dependence of the diamagnetic susceptibility. In that case the diamagnetic response exhibits a significant reduction for fields of the order of a few tens of $H_c(0)$ and is dominated by very low energy long wavelengths modes \cite{Skocpol75, Tinkham04}.

Summarizing we have observed a pronounced increase of the longitudinal spin-lattice relaxation in a narrow region close to the critical temperature of low $T_c$ thin superconducting films.
The peak is attributed to unexpected low frequency ($\sim$ kHz) critical fluctuations.  The strong field dependence  may explain  why these have not been detected with conventional NMR. Both the magnitude and the field dependence are not predicted in the existing theory of NMR relaxation near $T_c$.
Our observation also demonstrates the sensitivity of $\beta$-NMR to low frequency dynamics in superconductors. In contrast with conventional NMR, depth dependent $\beta$-NMR
can be carried out at very low frequencies  and thin films without  loss in signal strength.
The implantation range of the polarized nuclei makes them suitable to application in low dimensional systems where fluctuation effects are enhanced e.g.  heterostructures containing insulating and superconducting layers, where it can be stopped in the insulating environment to act as a sensitive probe of fluctuations in the nearby superconducting layer, or High-T$_c$ superconductors.

We thank T. Neupert, C. Mudry, A. Suter and B. M. Wojek, PSI, for useful discussions and M. D\"obeli, ETHZ, for performing RBS measurements on the film.
%
\bibliographystyle{prsty}

\end{document}